\documentclass[sigconf]{acmart}

\AtBeginDocument{%
  \providecommand\BibTeX{{%
    \normalfont B\kern-0.5em{\scshape i\kern-0.25em b}\kern-0.8em\TeX}}}

\setcopyright{acmcopyright}
\copyrightyear{2019}
\acmYear{2019}
\acmConference[MIG '19]{Motion, Interaction and Games}{October 28--30, 2019}{Newcastle upon Tyne, United Kingdom}
\acmBooktitle{Motion, Interaction and Games (MIG '19), October 28--30, 2019, Newcastle upon Tyne, United Kingdom}
\acmPrice{15.00}
\acmDOI{10.1145/3359566.3360076}
\acmISBN{978-1-4503-6994-7/19/10}

\usepackage{amssymb}
\usepackage{amsmath}
\usepackage{caption}
\usepackage{float}
\begin{document}

\title{A Robust Interactive Facial Animation Editing System}
\author{Elo\"{i}se Berson}

\affiliation{%
  \institution{Dynamixyz}
}
\affiliation{%
 \institution{CentraleSup\'{e}lec, CNRS, IETR, UMR 6164, F-35000}
 \city{Rennes}
 \country{France}}
\email{eloise.berson@dynamixyz.com}

\author{Catherine Soladi\'{e}}
\affiliation{%
        \institution{CentraleSup\'{e}lec, CNRS, IETR, UMR 6164, F-35000}
  \city{Rennes}
  \country{France}}
\email{catherine.soladie@centralesupelec.fr}

\author{Vincent Barrielle}
\affiliation{\institution{Dynamixyz}
}
\email{vincent.barrielle@dynamixyz.com}

\author{Nicolas Stoiber}
\affiliation{\institution{Dynamixyz}
}

\email{nicolas.stoiber@dynamixyz.com}

\renewcommand{\shortauthors}{Berson et al.}

\begin{abstract}
Over the past few years, the automatic generation of facial animation for virtual characters has garnered interest among the animation research and industry communities.
Recent research contributions leverage machine-learning approaches to enable impressive capabilities at generating plausible facial animation from audio and/or video signals.
However, these approaches do not address the problem of animation edition, meaning the need for correcting an unsatisfactory baseline animation or modifying the animation content itself.
In facial animation pipelines, the process of editing an existing animation is just as important and time-consuming as producing a baseline.
In this work, we propose a new learning-based approach to easily edit a facial animation from a set of intuitive control parameters.
To cope with high-frequency components in facial movements and preserve a temporal coherency in the animation, we use a resolution-preserving fully convolutional neural network that maps control parameters to blendshapes coefficients sequences.
We stack an additional resolution-preserving animation autoencoder after the regressor to ensure that the system outputs natural-looking animation.
The proposed system is robust and can handle coarse, exaggerated edits from non-specialist users. It also retains the high-frequency motion of the facial animation.
The training and the tests are performed on an extension of the B3D(AC)\^{}2 database \cite{fanelli_3-d_2010}, that we make available with this paper at \url{http://www.rennes.centralesupelec.fr/biwi3D}.

\end{abstract}

\ccsdesc[500]{Computing methodologies~Motion processing}
\ccsdesc[300]{Computing methodologies~Neural networks}
\ccsdesc[300]{Human-centered computing~Graphical user interfaces}

 \keywords{Facial animation, interactive motion edition, learning-based approach, dataset}

\begin{teaserfigure}
        \includegraphics[width=\textwidth]{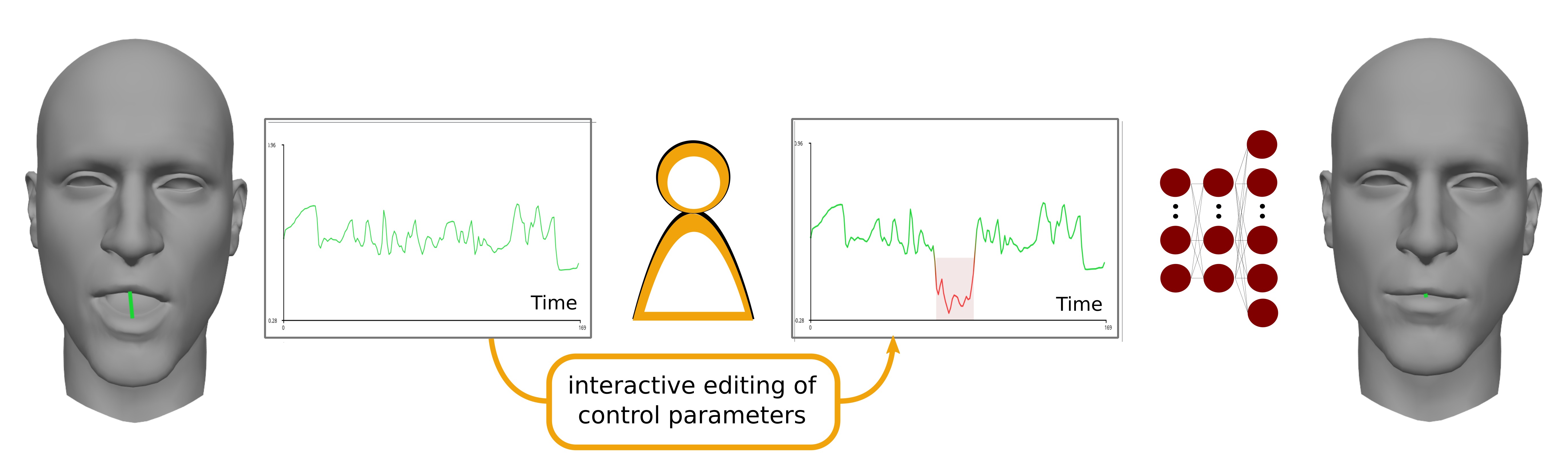}
  \caption{Editing pipeline. An interactive interface enables the user to easily edit meaningful control parameters that are automatically mapped to a realistic facial animation.}
  \Description{Editing pipeline.
   }
  \label{fig:teaser}
\end{teaserfigure}

\maketitle

\begin{figure}[t]
  \centering
  \includegraphics[width=.45\textwidth]{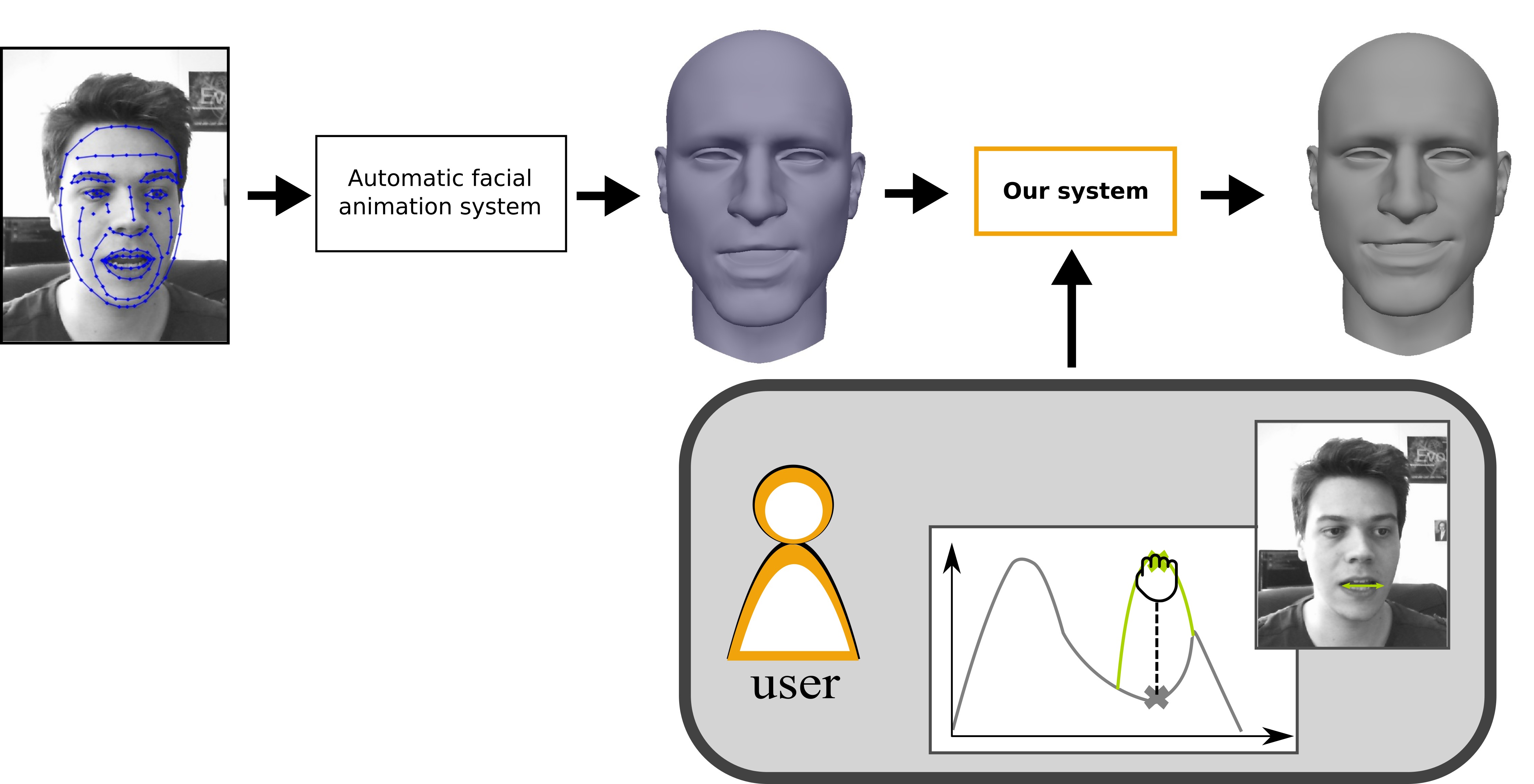}
  \caption{\
    \label{fig:overview}
    System overview. Our editing system allows a non-specialist user to easy and quickly interfere in the traditional facial animation pipeline to refine an animation.
  }
\Description{System overview: our system is designed to be integrated in the traditional facial animation pipeline.}
\end{figure}

\section{Introduction}

Producing a convincing facial animation for a virtual character is a tedious and time-consuming task which requires talent and experience, both rare resources.
Many works have been conducted in the field of animation research to come up with automatic facial animation generation systems that would accelerate this process.
Traditionally, automatic animation systems consist in an end-to-end pipeline, taking as input either audio~\cite{zhou_visemenet:_2018}, image or text~\cite{taylor_deep_2017},
and generating a full sequence of animation data (typically as a sequence of blendshapes coefficients).
Although impressive results have been reached, it is common to require human intervention to refine this baseline animation, either to correct mistakes or to make some adjustment to adapt the motion to another artistic intent.
These modifications are done by an animator trained in manipulating low level 3D animation parametrization such as blendshapes coefficients.
This edition pipeline implies technical and artistic skills, as well as a considerable amount of time to end up with a coherent and satisfactory final animation.

In this paper, we present an editing tool that allows non-specialist users to easily and quickly refine an existing animation. 
The main challenge in facial animation is to ensure the naturalness of the motion.
Indeed, the human are experts at observing faces and can detect even subtle implausible movements, notably the missing of lip contact when the mouth closes during a speech.
We develop a machine-learning-based approach that learn from a dataset, and trains to produce natural-looking animation from
a small set of input parameters.
By training on natural animation space-time patterns, our system learns to preserve the temporal coherency of the motion and ensure smooth and continuous animation.
As contributions such as Seol et al.~\cite{seol_spacetime_2012}, our system is designed to be efficiently integrated in the traditional facial animation pipeline as shown in Figure~\ref{fig:overview}.
However, unlike Seol et al.~\cite{seol_spacetime_2012} who focuses on producing an efficient system dedicated to a professional use, our goal is to provide an alternative solution for non-specialist users.
We specifically design our system to be robust to inadequate user edits, and handle exaggerate or conflicting inputs. 
Besides, instead of complicated facial control parameterizations we propose to use intuitive high-level control parameters as input to the system, such as specifying the distance between the lips over time.
The system runs at low latency, enabling us to create a graphical interface for users to interactively modify the output animation until getting a satisfying result.

One challenge when dealing with facial animation is to preserve the high-frequency patterns of the motion, as they are responsible for important communication cues (eye closures, lip contacts).
This is particularly true for learning-based solutions, that leverage large datasets of complex, possibly conflicting animation patterns~\cite{holden_deep_2016}.
Among the shortcomings of these solutions is the ability to preserve the different frequency components of the animation and to adapt the behavior of the system to inconsistent inputs.
In this work, we define an architecture based on a fully convolutional network with skip layers designed specifically to preserve high-frequency components. 
Besides, we aim at a system that is resilient to coarse editing by non-specialist users.
To that end, we train an additional denoising autoencoder that we stack at the end of the network to ensure a natural-looking final animation output.

In order to be a suitable solution for non-expert users to create powerful facial animation pipelines, an editing tool has to meet the following requirements: 
1. Usability: a user should be able to personalize a 3D animation without advanced animation skills.
The number of control parameters should be small, and those should be meaningful and easy to manipulate.
The system should run fast enough to enable interactive editing. 
The user can then iteratively modify its animation, either by editing a few frames or by imposing full-sequence constraints, until a satisfying result is produced (see Section~\ref{sec:qualitative_res}).
2. Plausibility: the complex space-time patterns of human facial motion should be respected (see Section~\ref{sec:comp_with_human_motion}). 
In particular, high-frequency facial movements should be present. 
3. Robustness: the final animation should remain plausible regardless of the user modifications (see Section~\ref{sec:behavior_analysis}).
4. Subject and content independent: any type and style of facial animation should be able to be edited (see Section~\ref{sec:test_voca}).
In the following sections, we describe a learning-based editing system that addresses all these requirements.

Our machine-learning system relies on a dataset of facial animation sequences.
To train the full system, we worked on modifying and extending the existing 3D facial animation dataset B3D(AC)\^{}2 database~\cite{fanelli_3-d_2010}.
We will release the extended dataset for reproducibility of our results.

Our contributions are:
\begin{itemize}
        \item A new facial animation editing system based on convolutional neural networks, which enables to quickly edit a temporal talking facial animation with few intuitive 
              control parameters.
              Based on a time resolution-preserving architecture, our system is capable of generating complex and plausible facial motion pattern. 
              The proposed framework features a regressor designed to map low dimensional control parameters to blendshapes coefficients sequences.
              It is followed by an autoencoder meant to ensure the naturalness of the outputted animation sequences.\\ 
        \item A robust solution dedicated to non-specialist users that is resilient to implausible inputs constraints. 
                We use a denoising training strategy to improve the reliability of our system.
                The originality comes from the indirect noisy inputs used to train the stacked autoencoder, and an additional loss term encouraging mouth
                closure preservations during talking facial animations. \\
        \item The release of our enhanced 3D audiovisual animation database from Fanelli et al.~\cite{fanelli_3-d_2010} with notably a parameterization of the animations with the widely-used blendshapes formalism.
                With the use of a professional software we added 2D eyelids and mouth annotations and improved the overall quality of the animation and the depiction of cues such as eyelids and lip contacts.
\end{itemize}

\section{Related Work}

\begin{figure*}[t!]
  \centering
  \includegraphics[width=.95\textwidth]{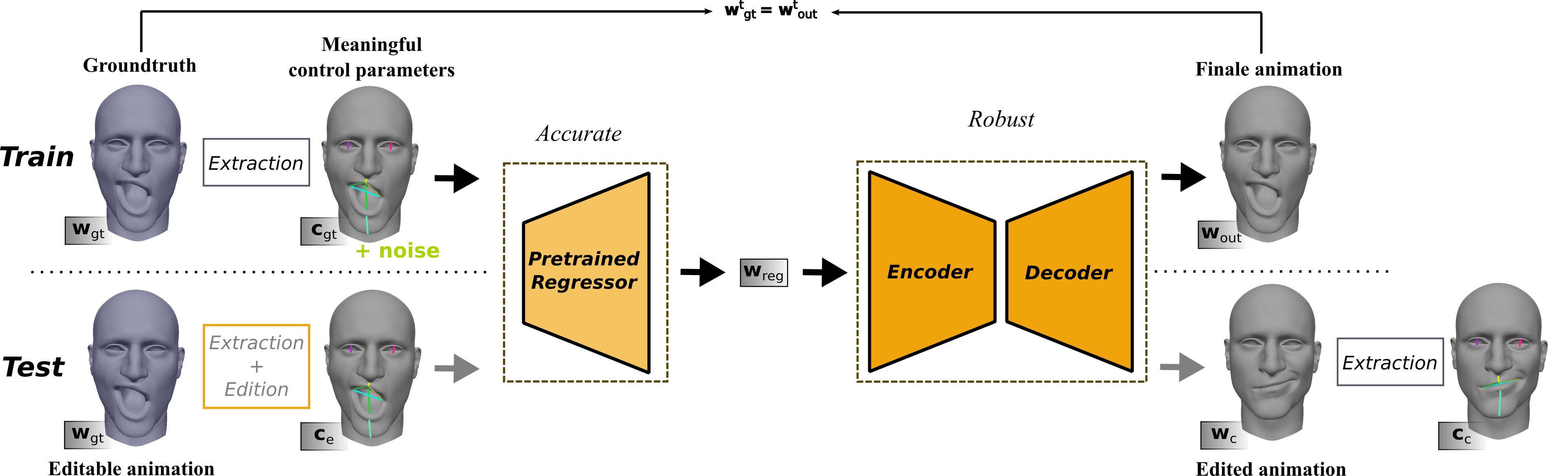}
  \caption{\
    \label{fig:pipeline}
    System description. (Top) At train time, fixing the parameters of the regressor, the autoencoder learns to reconstruct the initial blendshape weights from the noisy meaningful control parameters.
    (Bottom)  At test time, the edited control parameters lead to an accurate blendshapes weights sequence thank to the regressor. The stacked autoencoder allows inaccurate edition ensuring a realistic edited animation.
  }
  \Description{System description. (Top) At train time, fixing the parameters of the regressor, the autoencoder learns to reconstruct the initial blendshape weights from the noisy meaningful control parameters.
  (Bottom)  At test time, the edited control parameters lead to an accurate blendshapes weights sequence thank to the regressor. The stacked autoencoder allows inaccurate edition ensuring a realistic edited animation.}
\end{figure*}

In the professional world, animation edition is done by directly manipulating the temporal curves of complex facial parameterizations (blendshapes coefficients being an industry-standard parametrization).
Hence, traditional animation production requires animators with technical and artistic skills, experience and remains time-consuming even for those.
Previous works have addressed this problem by providing efficient animation editing systems either based on geometry-driven approaches or data-driven approaches.

\subsection{Geometric animation edition}

Early facial expression manipulation approaches are based on key frames edition.
The key frames can be made of linear combinations of face meshes coming from a pre-captured database~\cite{joshi_learning_2003, zhang_geometry-driven_2003, zhang_spacetime_2004, chi_interactive_2017}.
The goal is to find the blending weights corresponding to user constraints.
In these approaches, the user manipulates 2D control points which are either image features~\cite{zhang_geometry-driven_2003}, motion markers~\cite{joshi_learning_2003}
or the 2D projection of 3D vertices~\cite{zhang_spacetime_2004, chi_interactive_2017}.
Other works consider the key frame editing problem as solving the 3D vertices position of the edited mesh.
To reduce the dimension of the facial model, Lau et al.~\cite{lau_face_2009} used PCA to obtain a subspace representation.
Then, they derived the 3D face vertices positions from strokes, points or curves constraints drawn by the users on a 2D screen.
An alternative to PCA to obtain a semantically meaningful data representation is the Independent Component Analysis (ICA)~\cite{cao_unsupervised_2003, ma_style_2009}.
This parametrization gives the possibility to distinguish between editing facial emotional components and speech-related components.
More recent works develop editing systems that can be easily integrated in the animation pipeline.
The animation editing problem consists in finding the underlying blendshapes coefficients.
The users directly manipulate the vertices of the mesh~\cite{lewis_direct_2010, anjyo_practical_2012, tena_interactive_2011} or draw 2D strokes on a screen~\cite{cetinaslan_sketch-based_2015}.
The number of users constraints is generally smaller than the blendshapes model parameters, the optimization problem is thus regularized through different criteria: by constraining the value of the blendshapes coefficients~\cite{lewis_direct_2010},
by using a statistical model~\cite{anjyo_practical_2012, weise_realtime_2011}, by constructing an orthogonal blendshapes model~\cite{li_orthogonal-blendshape-based_2008}, by using geometric constraints~\cite{ribera_facial_2017} or by adding face areas boundary constraints~\cite{tena_interactive_2011}.
To improve the applicability of the edited method, most of the previous works segment the face into hierarchical regions a priori~\cite{joshi_learning_2003, zhang_geometry-driven_2003, li_orthogonal-blendshape-based_2008, ma_style_2009},
or a posteriori using an influence map for each control points~\cite{zhang_spacetime_2004, chi_interactive_2017}, or else using the ICA transform for decorrelation~\cite{cao_unsupervised_2003}.\\
These methods are mainly frame-based and most of them do not consider the temporal consistency of movements.

To overcome that limitation, there are works dealing with the dynamic nature of animation data, instead of performing edition on static expressions.
Li and Deng~\cite{li_orthogonal-blendshape-based_2008} propose sequence edition by fitting a Catmull-Rom spline on the edited blendshapes weight sequences.
This technique does not necessarily preserve the naturalness of the motion, as nothing encourages the motion to be physically correct.
Inspired from space-time constraints body motion systems~\cite{gleicher_motion_1997}, Ma and colleagues~\cite{ma_style_2009} create a style learning editing framework.
The editing of style is applied to similar frames in the sequence.
While it is an efficient solution to reduce time spent in the animation editing process, this solution does not ensure temporal coherency.
Applying the same edit to similar frames with a different context leads to inconsistent motion.
Seol et al.~\cite{seol_spacetime_2012} and Akhter et al.~\cite{akhter_bilinear_2012} propose a temporal solution to propagate the edits across the surrounding frames by solving a movement matching equation or by using a
spatiotemporal bilinear model.
Although, these methods provide smooth results, their temporal resolution depends on hyperparameters that need to be manually adjusted rendering the editing task more difficult to tune.
Moreover, the system of Seol et al.~\cite{seol_spacetime_2012} is not robust to inconsistent users edits.
For example, in the case of exaggerated user constraints, this method generates implausible animations.
As we target non-specialist users, our system needs to ensure the final motion to be realistic.

\subsection{Data-based animation edition}
In contrast to keyframe-based geometric editing methods, space-time methods consider the manipulation of entire temporal motion data-blocks.
An effective technique to perform temporal coherent edition and generation is motion graphs~\cite{kovar_motion_2002, zhang_spacetime_2004}.
This technique consists in building a graph where nodes encode static poses or short-term motion blocks. The graph can be navigated to recreate plausible animation sequence.
The edges between nodes encode the likelihood of the transition between those two blocks being plausible, so realistic animation reconstruction consist in finding paths of minimal cost in the graph.
While motion graph is a relevant technique for our purpose, it imposes high memory usage as it requires retaining the whole graph for inference.
Moreover, a balance has to be achieved between expressivity, which can be obtained by a graph with a large number of connections,
and physical consistency, which is better enforced with a sparser graph featuring only consistent transitions.

More recent works in the line of data-based method have adopted new models for space-time human motion editing systems.
The first one to propose a fully learning-based human motion editing system is the seminal work of Holden et al.~\cite{holden_deep_2016}.
They map high level control parameters to a learned body motion manifold presented earlier by the same authors~\cite{holden_learning_2015}.
Navigating this manifold of body motion allows to easily alter and control body animations, while preserving their plausibility.
Recently, Habibie and colleagues~\cite{habibie_recurrent_2017}, as well as Martinez and coworkers~\cite{martinez_human_2017}, designed state-of-the art dynamic motion modeling systems, demonstrating the high potential of learning-based approach in human motion manipulation.

Closest to our work, tackling the same challenge of editing an animation using simple high level parameters, is the work of Holden et al.~\cite{holden_deep_2016}.
Unlike body motion however, facial motions have a lot of high-frequency temporal components such as blinking, lips synchronization, and mouth closures.
Although the system of~\cite{holden_deep_2016} demonstrated impressive results on body motion, we found that their architecture is not particularly suited to this particular aspect of facial animation.
Using it in our scenario leads to over-smoothed, unappealing facial animations, which we illustrate in Section~\ref{sec:experiments}.
We therefore adapt this approach for the purpose of facial motion, and tackle the high-frequency issue using a resolution-preserving neural network.
Our work is based on a one dimensional fully convolutional network inspired from Ronnenberger and colleagues~\cite{navab_u-net:_2015}, with skip connections
between the down-sampling and the up-sampling parts in order to preserve high-frequency details.
To the best of our knowledge, we are the first to study a temporal editing system based on a resolution-preserving neural network.

The remainder of the paper is organized as follows.
In Section~\ref{sec:dataset}, we will present the dataset work we have conducted, which enabled us to train and test our model.
We then describe our model in Section~\ref{sec:method}. We focus particularly on the benefit of the added autoencoder and the specific way of training it.
We compared our system with related works in Section~\ref{sec:experiments} and conducted several experiments highlighting the performance and benefits of our architecture.
Finally, we demonstrate the usability of our framework in a realistic animation production pipeline.

\section{Dataset preparation}

\label{sec:dataset}
For our experiments, we use the 3D Audio-Visual Corpus of Affective Communication \cite{fanelli_3-d_2010}, which contains 3D scans and RGB images of 14 actors reciting 40 sentences with and without emotion.
However, the quality of scan data prevents us from having a good depiction of subtle mouth closures and blinking. Those are however crucial to verbal and non-verbal communicational cues that facial animation convey.
Besides, in order to integrate this tool to the traditional facial animation pipeline, we want facial expressions to be encoded through the standard blendshapes parametrization, which will be used as input to our system.

We address the above issues by fitting a common deformable template, with a sparser mesh, to the neutral geometry of each actor. 
Then, we transfer a blendshape model onto the aligned deformable template. \\ 

The alignment consists of three stages. 
First, we align a 3D morphable model~\cite{blanz_morphable_1999} with the neutral mesh of each actor using a non-rigid ICP algorithm, optimizing the pose and the identity coefficients.
Inspired by Li et al.~\cite{li_realtime_2013}, we improve the quality of the alignment around the mouth and eyelids using 2D image landmarks information for each frame, obtained with a commercial face tracking software~\cite{dynamixyz_performer_2019}, still optimizing for pose and identity coefficients.
We further refine the process by optimizing the vertices directly, through a non-rigid ICP with Laplacian prior~\cite{li_robust_2009}.
Then, using deformation transfer~\cite{sumner_deformation_2004}, we transfer a pre-existing blendshape model sharing the topology of the morphable model onto the deformed mesh.
At this point, we obtain a subject-specific blendshapes model for the 14 actors.
Finally, we derive our final dataset of 29 blendshapes animation weights by fitting the model on the tracked 2D landmarks in each frame of each actors' video performance using~\cite{dynamixyz_performer_2019}.

\section{System Description}

\label{sec:method}

In this section, we describe our facial animation editing system in more detail. 
First, we justify the choice of the control parameters which constitutes the input of our system.
Then, we elaborate on the structure of the neural network that forms the heart of our system.
The network is composed of two parts.
The first part is a regressor which maps high-level inputs to a blendshapes weights sequences.
The second one is a stacked autoencoder that cleans the blendshapes weights sequence to ensure a realistic final animation.
Both are fully convolutional, and operate on space-time signals, meaning they perform temporal convolutions on a time window of their input parameters.

\subsection{Meaningful high-level control parameters}
\label{sec:cp}
\begin{figure}[h]
  \centering
  \includegraphics[width=.15\textwidth]{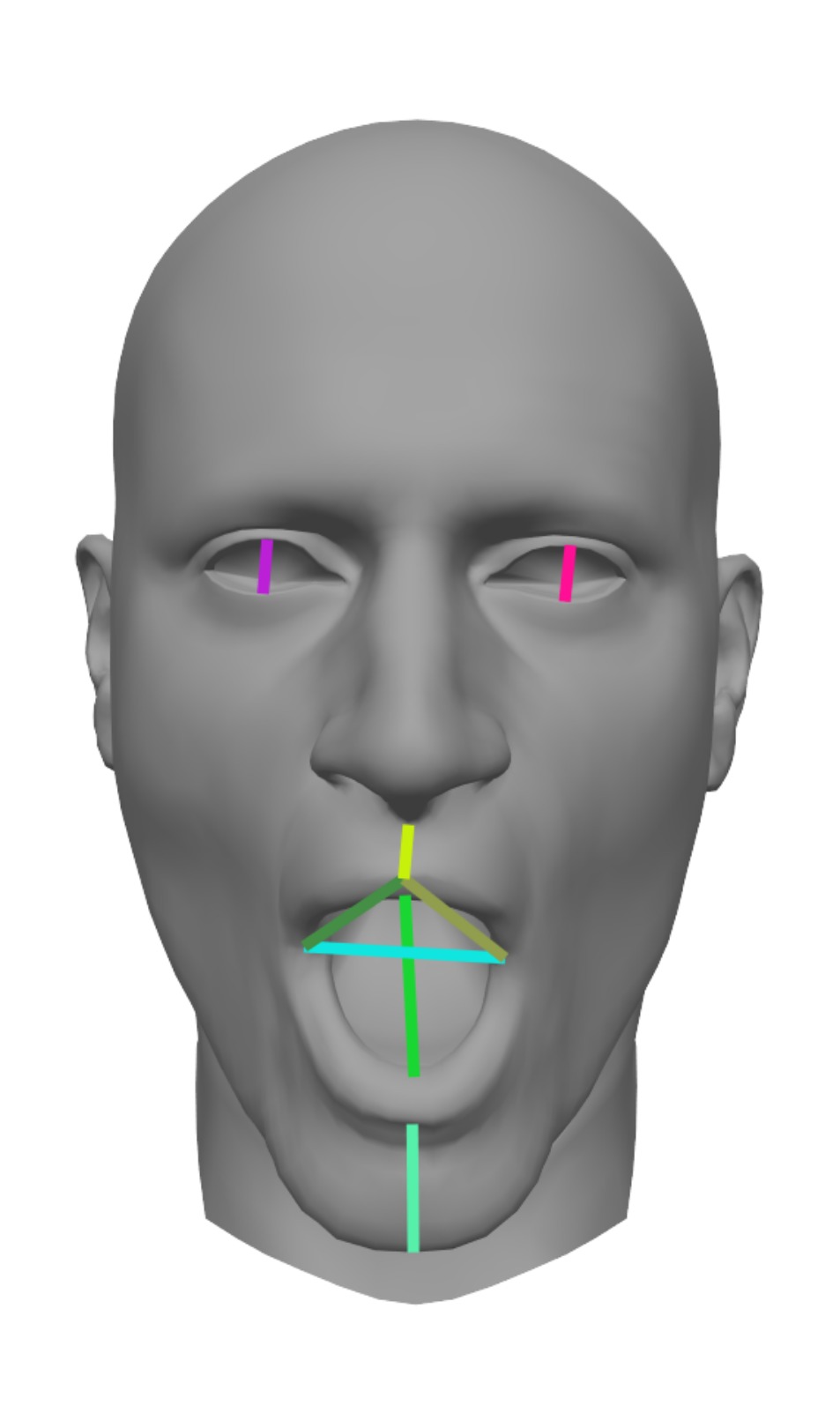}
  \caption{\
    \label{fig:control_para}
    Eight meaningful control parameters extracted from the mesh.
  }
  \Description{Eight meaningful control parameters extracted from the mesh.}
\end{figure}

We aim at a system that takes intuitive high-level parameters as input, for users to easily translate their desired modifications into animation.
Particularly important in facial animation is the rendition of speech, so we want the control parameters to be able to specify all plausible mouth shapes that occur during natural speech.
These parameters have to be simple and meaningful to be intuitively manipulated by non-professional users.
Thus, we choose eight inter-vertex distances shown in Figure~\ref{fig:control_para} as our control parameters.
The horizontal and vertical inner-lips distances as well as the eyelids distances determine the state of mouth/eye closure, two important expressive cues.
To enable editing the emotional expressiveness of the animation such modifying the smile intensity, we add the distance between the upper-lip center and the mouth corners.
The lips protrusion, activated by pronouncing palate sounds such as "sh or ch" or doing a kiss shape are manipulated with the distances between the nose bridge and the upper-lip center and 
between the chin and the bottom-lip center. 
We found this to be a rather minimal set for our approach. Less parameters would result in ambiguous specifications for face shape, leading to a noisy regressor output.
In this work we always measure those distances on a blendshapes-based character with fixed morphology. This ensures that the distance patterns we extract from the dataset's animations of section~\ref{sec:dataset} are actor-independent.

While our network can learn full-face motion patterns, we found that generalization of the results is improved if we split the facial controls in three groups that exhibit low motion correlation with each other in the database: lower-face, upper-face and eyelids.
An independent network will be trained for each group, with its own relevant high-level control parameters as input, and appropriate blendshapes coefficients as output.
This splitting of the face is common in previous research works and practical applications~\cite{joshi_learning_2003, zhang_spacetime_2004}.

\subsection{Regression from low dimension control parameters to blendshapes weights}

Motivated by the observation that facial animation is composed of high-frequency features, we moved away from previous motion-modeling network architectures and built a resolution-preserving neural network to regress the control parameters ($\mathbf{c}_{gt}$) to blendshapes weights $\mathbf{w}_{reg}$ as shown in Figure~\ref{fig:pipeline}.
The value for control parameters have been calculated on a fixed morphology character, animated with the blendshapes weights ($\mathbf{w}_{gt}$) extracted from the database.

\begin{figure}[h]
  \centering
  \includegraphics[width=.45\textwidth]{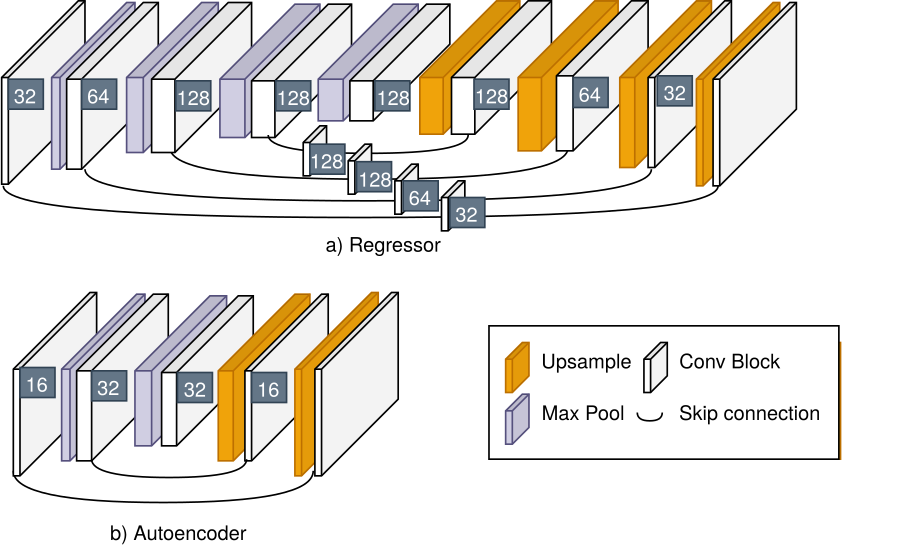}
  \caption{\
    \label{fig:regressor}
    Architecture of the regressor (a) and the autoencoder (b)
  }
  \Description{Architecture of the regressor and the autoencoder}
\end{figure}

The regressor is a fully one-dimensional convolutional neural network with skip layers, a structure sometimes loosely described as U-net.
Its architecture is depicted in Figure~\ref{fig:regressor}a.
We use one-dimensional max-pooling layers and up-sampling layers to respectively down-sample and up-sample the temporal dimension.
Each convolutional block in Figure~\ref{fig:regressor} is composed of a batch normalization layer, a convolutional layer and the elu activation function~\cite{clevert_fast_2015}.
As input to the regressor we use a time-window of 64 frames. We extract those windows from complete sequences with a time-overlap ratio of 0.75.
As preprocessing, we subtract the mean controller values calculated on the whole trainset. 
All the filters in the network have a size of 3.
Our loss function is composed of two terms~\cite{holden_deep_2016}: the mean square error (MSE) between the $\mathbf{w}_{gt}$ and $\mathbf{w}_{reg}$, $\mathcal{L}_{MSE}$, and a L2 regularization on the weights $\beta * \mathcal{L}_{reg}$.
We set the tradeoff parameter $\beta$ equals to 1. 
We employ the Adam optimizer for training with a batch size of 128 and an initial learning rate of 0.001 with a decay ratio of 0.95 every five consecutive epochs with no validation loss improvement.

We use sequences from 13 subjects of the dataset to train our network.
This amounts to around 85 minutes of facial animation, which we split into a training set and a validation with a 0.95 ratio.
The final state of the network we conserve is the epoch  the lowest validation loss.

\subsection{Autoencoder for ensuring the naturalness of the animation}

Our network features an animation autoencoder whose role is to clean-up the output of the regressor.
Our regressor is a rather straightforward mapping network, so it will faithfully transcribe any user command, easily extrapolating to cases of unrealistic facial animation.
However, we opted for a robust system, which would unsure staying in a realistic animation space no matter the user input.  
The added autoencoder serves that purpose.
Its architecture is depicted in Figure~\ref{fig:regressor}b.

Ensuring that the network produces realistic animation is due to both the presence of the autoencoder and to the following denoising training strategy.
Training autoencoders as denoisers -meaning feeding them with noised inputs and clean outputs- is common practice, but we found that the resulting autoencoder is very dependent on the noise characteristics.
In our case, since the noise is supposed to mimic unrealistic user inputs, we found it difficult to find a good noise model.
Instead we chose to train the whole end-to-end system as a denoiser, while keeping the regressor weights constant (except the statistics of the batch normalization layers) and optimize the autoencoder's weights to reconstruct $\mathbf{w}_{gt}$.
In practice, we modify around 20\% of the control parameter inputs of the regressor $\mathbf{c}_{gt}$ (see Figure~\ref{fig:pipeline}) with salt-and-pepper noise. 
We found that this creates noisy animation patterns for the autoencoder to train to clean-up that are closer to what the system would encounter in a real runtime scenario.
For the autoencoder to conserve the high-frequency features of the regressed output we use a convolutional architecture similar to that of the regressor (Figure~\ref{fig:regressor}).

The blendshapes parameterization is not the most representative of the importance of each movement they encode. Movements such as mouth openings/closures carry more expressive and communicational weight than others such as nose movements.
The loss that our network learns to minimize should reflect this aspect.
To the MSE loss on all blendshapes coefficients we therefore add a loss on the difference between some intervertices distances on the model animated with $\mathbf{w}_{gt}$ and $\mathbf{w}_{out}$.

\begin{equation}
        \mathcal{L} = \mathcal{L}_{MSE} + \alpha * \mathcal{L}_{distance} 
\end{equation}

Typically, $\mathcal{L}_{distance}$ measures the distances between the lips, and between the eyelids.
This term helps ensuring an accurate mouth closure during a talking facial animation~\cite{ma_real-time_2018}.
For our experiments, the parameter $\alpha$ is set to 1.
Training the model takes less than 2 hours on a NVIDIA GeForce GTX 1070 GPU.

\section{Experiments \& results}
\label{sec:experiments}
In this section, we present experimental results of our facial animation editing system.
First, we evaluate our system by comparing its integrity to the recent related work of \cite{holden_deep_2016}, which addresses a similar set of requirements, albeit for body animation applications. We retained their system's architecture, adapting it for our specific inputs and outputs. The discrepancy between quantitative measures and qualitative look of the animations lead us to use special metrics for a more complete comparison (Section~\ref{sec:comp_with_human_motion}). 
This comparison confirms the suitability of the proposed neural network system for the purpose of facial animation, as well as its capacity to create plausible facial animation preserving the complex dynamic of the facial movements. 

To assess the data-dependency and reproducibility of our system, we apply it on a different recently released database (see Section~\ref{sec:test_voca}) and measure quantitative performance.
In Section~\ref{sec:behavior_analysis}, we study the robustness of our system to implausible user constraints, and analyze the role of the system's components.
Finally, as our system runs with low latency, we demonstrate in Section~\ref{sec:qualitative_res} its potential as an interactive animation tool by showing examples of edition performed on animations resulting of facial tracking.

\subsection{Comparison with state-of-the-art approach}
Our system is designed for animation edition and control, but it will only be useful if its architecture can handle and represent sufficiently varied facial motion. Of particular interest is the ability to preserve the high-frequency components of facial animation, which are important for human communication.
In practice, we evaluate how close the generated animation $\mathbf{w}_{c}$ is to ground-truth $\mathbf{w}_{gt}$ when the edited control parameters $\mathbf{c}_{e}$ are kept unchanged, equal to $\mathbf{c}_{gt}$ (see Figure~\ref{fig:pipeline}).
We evaluate this metric on the whole database using the leave-one-subject-out strategy.

\label{sec:comp_with_human_motion}
To our knowledge, there is no work directly addressing the problem of high-level, temporal consistent manipulation of facial animation.
In the broader field of animation research, Holden et al.~\cite{holden_deep_2016} set to tackle a similar set of goals for body animation editing and control. Part of their system is valid for facial animation and can be adapted to our inputs and outputs.

To represent their approach, we first learn a time-convolutional autoencoder with one layer to encode the sequence animation into a latent space and one layer to decode.
Then, we learn a fully convolutional network to regress the control parameters to this latent space (see \cite{holden_deep_2016} for the details).
The regressor is built with only 2 layers as it appeared to give better results in our case.
To get a fair comparison, we train one such system by face area, similar to ours (see Section~\ref{sec:cp}).

We evaluate the different systems by minimizing the mean square error (MSE) between the input and the output blendshapes weights sequences.
For our experiments, we use the regressor with the lowest MSE because the role of the regressor is to accurately regressed the control parameters to the blendshapes weights.

\begin{table}
\caption{
        \label{tab:quantitative_res}
Quantitative comparison between the regressor and the full system on the test set.} 
        \centering
\begin{tabular}{r|cc}
    \toprule
        & MSE  (lower face)& MSE  (eyelids)\\ 
    \midrule
        Regressor only & 0.0028 &  0.0064  \\
        Holden et al.~\cite{holden_deep_2016} & 0.004 & 0.009 \\
        Our system & 0.0082 & 0.0086 \\
    \bottomrule
\end{tabular}
\end{table}

\begin{figure}[h]
  \centering
  \includegraphics[width=.40\textwidth]{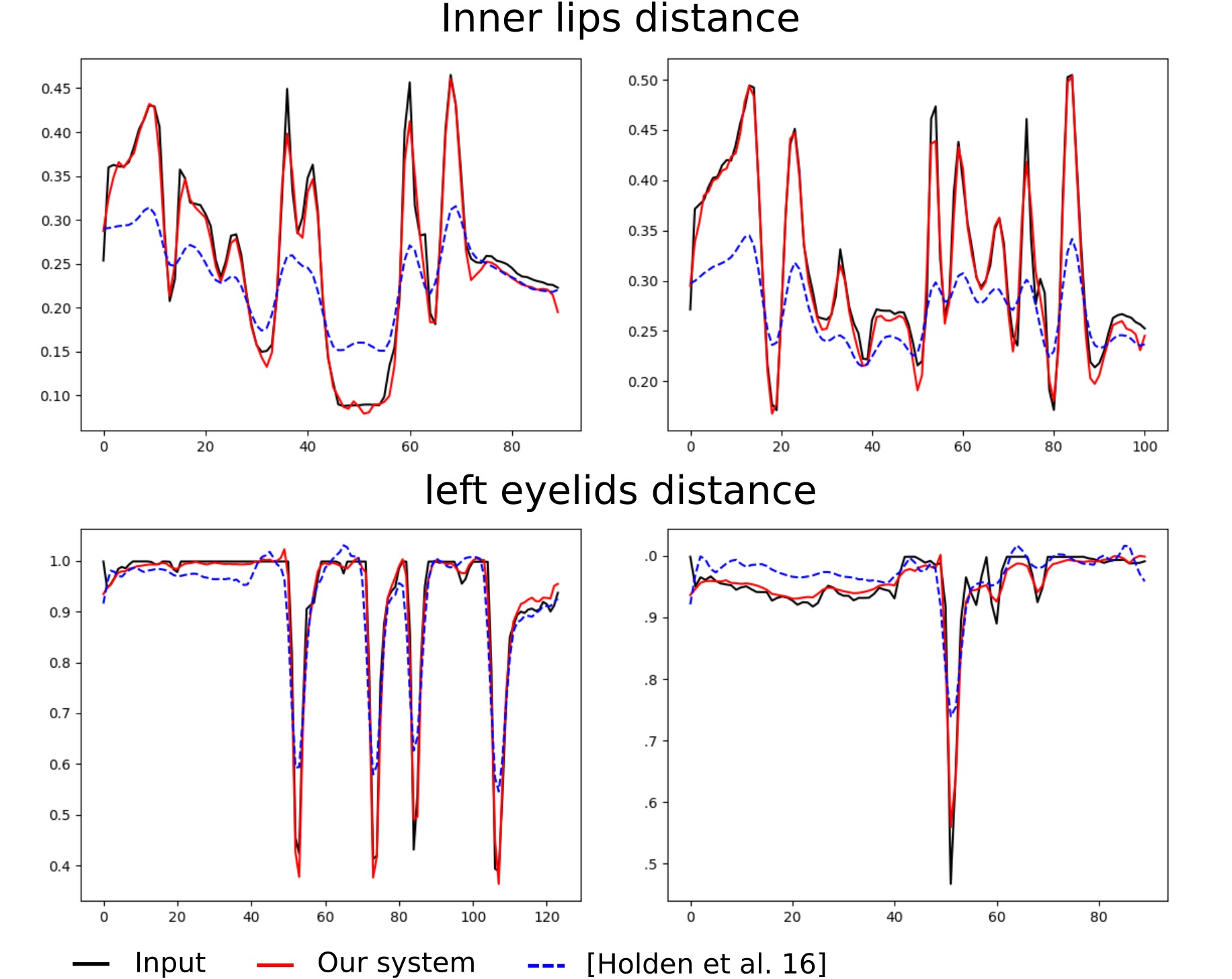}
  \caption{\
          \label{fig:comparison_holden}
          Comparison with Holden et al.~\cite{holden_deep_2016}:
          Curves of inner lips distance for different sequences. The body motion system \cite{holden_deep_2016} smoothes the output signal loosing the high frequency components.
  }
  \Description{Comparison with Holden et al.~\cite{holden_deep_2016}:
  Curves of inner lips distance for different sequences. The body motion system \cite{holden_deep_2016} smoothes the output signal loosing the high frequency components.}
\end{figure}

\begin{figure}[h]
  \centering
  \includegraphics[width=.45\textwidth]{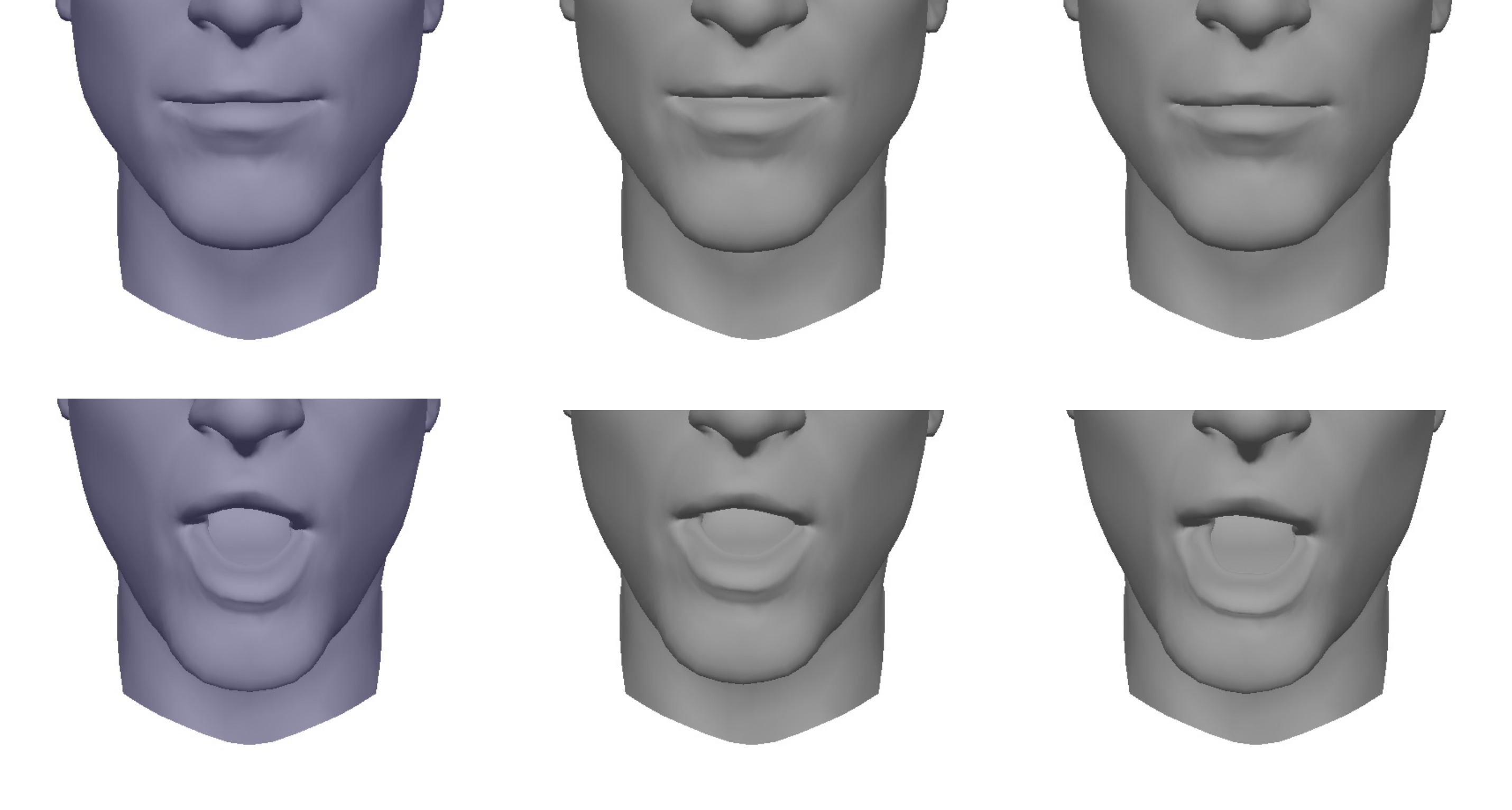}
  \caption{\
          \label{fig:image_holden}
          The groundtruth (left).
          Compare to \cite{holden_deep_2016} (middle), our system (right) is able to generate an animation
         which faithfully respects the input mouth movements and its amplitude.}
         \Description{
          Comparison to \cite{holden_deep_2016}: our system is able to generate an animation
         which faithfully respects the input mouth movements and its amplitude.}
\end{figure}

Interestingly, Table~\ref{tab:quantitative_res} shows that Holden et al.~\cite{holden_deep_2016} performs better than our complete system in term of MSE.
However, by looking at the temporal curves of inner lips distance derived from $\mathbf{c}_{gt}$ and ($\mathbf{c}_{c}$),
we realize that their system smooths the motion signal and shows consequent loss of high-frequency components of the mouth and the eyes (Figure~\ref{fig:comparison_holden}).
While the reconstruction MSE is lower, the corresponding animation is qualitatively less appealing as it misses the key high-frequency communicational cues on the mouth and eyelids.
Note that this behavior was probably less an issue in their original application on body animation, as high-frequency components carry less semantic weight in that case as it does for facial motion.
In Figure~\ref{fig:image_holden}, we display two frames extracted from sequences created from the same $\mathbf{c}_{gt}$ with the system of Holden et al.~\cite{holden_deep_2016} and our system.
We can see that, while our system produces an animation with faithful mouth openings and closures, the animation resulting of their system misses these cues due to the smoothing nature of their architecture.
Examples of animations using both systems
are shown in the supplementary video.

\begin{figure}[h]
  \centering
  \includegraphics[width=.45\textwidth]{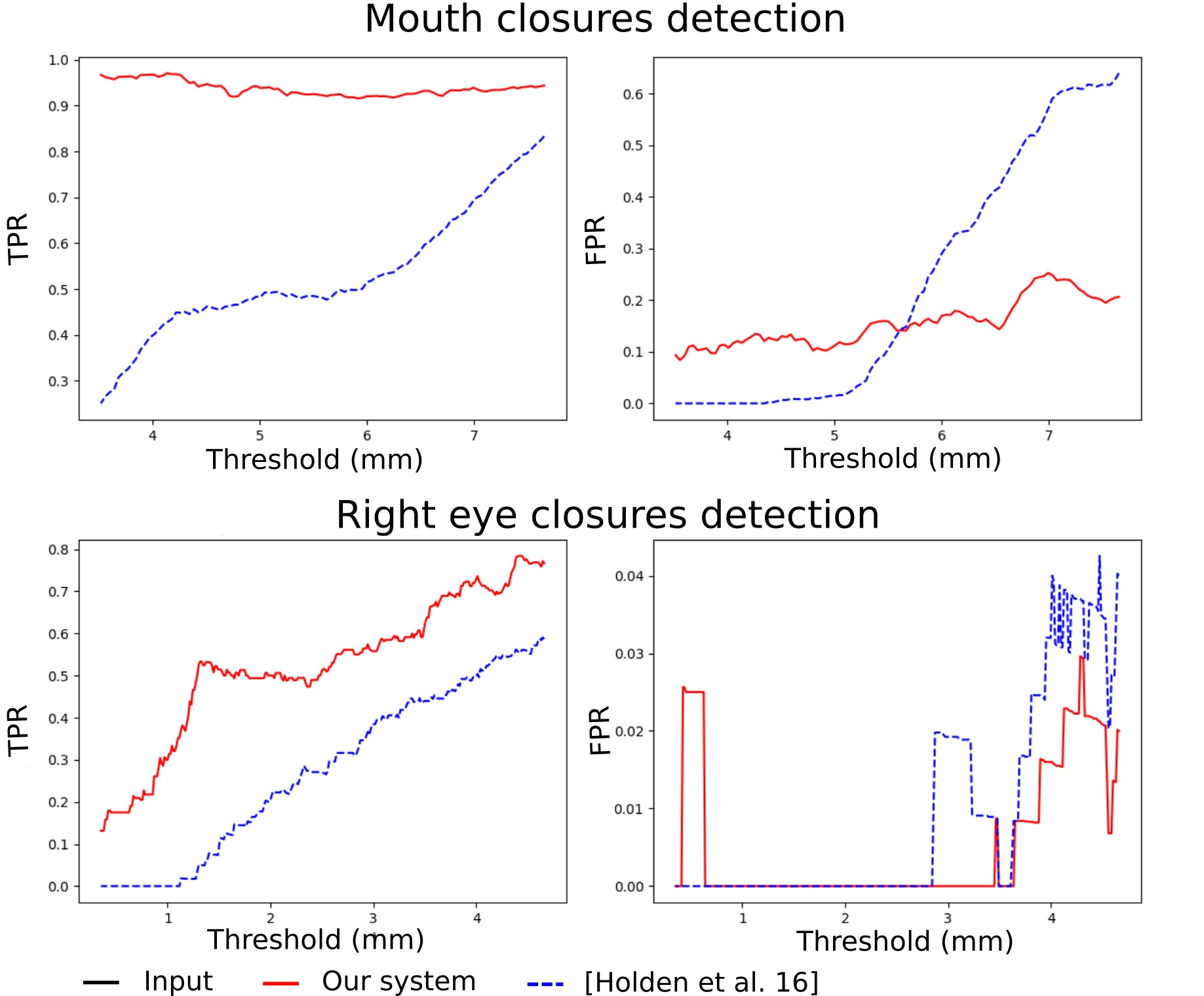}
  \caption{\
          \label{fig:own_metrics_holden}
          Comparison with \cite{holden_deep_2016}:
          Curves of the TPR and the FPR of the mouth and eyes closures on the testset.}
        \Description{
          Comparison with \cite{holden_deep_2016}:
          Curves of the TPR and the FPR of the mouth and eyes closures on the testset.}
\end{figure}

For a more representative quantitative comparison between our system and Holden et al.~\cite{holden_deep_2016}, we propose using a metric that highlights the capacity to accurately retain facial animation cues such as mouth contacts, closures and eye blinks.
To our knowledge, there is no agreed-upon metric in the community for such semantic facial cues, so we suggest measuring a true positive rate (TPR), i.e. ratio of true positive mouth- (respectively eyelid-) closures to the number of actual mouth- (eyelid-) closures, and the false positive ratio (FPR) defined
as the ratio of false positive mouth- (eyelid-) closures to the actual mouth- (eyelid-) closures. 
The TPR measures the capacity of the system to accurately preserve the desired mouth- and eye-related conversational cues.
The FPR controls that the system does not hallucinate undesired such movements.
On Figure~\ref{fig:own_metrics_holden}, we plot the TPR and the FPR for the mouth and right eyelid closures according to the threshold of detection.
We can see that for lower thresholds, only our system creates consistent mouth/eyes closures as its TPR is always the highest. 
The system of Holden et al.~\cite{holden_deep_2016} is not capable of producing eyes closures so its FPR is zero for lower thresholds.
Meanwhile, we control that our system does not hallucinate motion as its FPR remains low.

An interesting feature of data-based motion models is the ability to model immobility, that we observe here on the first curve plotting the inner lips distance in Figure~\ref{fig:comparison_holden}.
Between the $40^{th}$ frame and the $60^{th}$ frame, we can observe that our system can cope with no inner lips movements for multiple consecutive frames.

\subsection{Data dependency: transfers on another database}
\label{sec:test_voca}

As with all data-based approach, it is important to know how the approach depends on the size and content of the dataset.
Thus, we test the validity of our model (trained with the B3D(AC)\^{}2 dataset) on the recently released Vocaset database \cite{cudeiro_capture_2019}.
This dataset is composed of sequences of 12 subjects speaking sentences from the TIMIT corpus.
We use the same processing pipeline to get the blendshapes coefficients sequence as in Section~\ref{sec:dataset} except that we do not use 2D information.
We downsample the frame rate to 25 fps to match the frame rate of our dataset B3D(AC)\^{}2. 

As shown in Table~\ref{tab:sys_voca}, our system trained with only the trainset of the B3D(AC)\^{}2 dataset and applied to the whole Vocaset gives a comparable MSE (0.004) as a one trained with both the Vocaset and B3D(AC)\^{}2 dataset (0.003).
The Vocaset content is less diversified, that is why the results obtained using only this dataset are the lowest.
Indeed, there is no emotional sequence in this dataset unlike in the B3D(AC)\^{}2 dataset which is one-half composed with emotional sequence.
In such sequences, the amplitude of the movements is generally higher compared to neutral sequences.
So, at test time, it is easier for a system trained with emotional content to render neutral speech content than in the reverse order.
We can see on the supplementary material that our system is suitable to model any new subjects in the Vocaset.

\begin{table}
\caption{
        \label{tab:sys_voca}
Quantitative results of our system trained with the trainset of the B3D(AC)\^{}2 dataset.} 
        \centering
\begin{tabular}{r | r | c c c}
    \toprule
        Trainset & Testset &  MSE (mouth) & TPR & FPR \\
    \midrule
        Vocaset & Vocaset & 0.038 & 0.87 & 0.06 \\
        Vocaset & B3D(AC)\^{}2 & 0.05 & 0.81 & 0.38 \\
        B3D(AC)\^{}2 & Vocaset & \textbf{0.004} & \textbf{0.98} & 0.22   \\
        B3D(AC)\^{}2 & B3D(AC)\^{}2 & 0.008 & 0.98 & 0.22   \\
        Both & Vocaset & \textbf{0.003} &  0.95 & 0.22   \\
        Both & B3D(AC)\^{}2 & 0.01 & 0.95 & 0.35   \\

    \bottomrule
\end{tabular}
\end{table}

\subsection{System Robutness: necessity of the autoencoder}
\label{sec:behavior_analysis}
\begin{figure}[h]
  \centering
  \includegraphics[width=.45\textwidth]{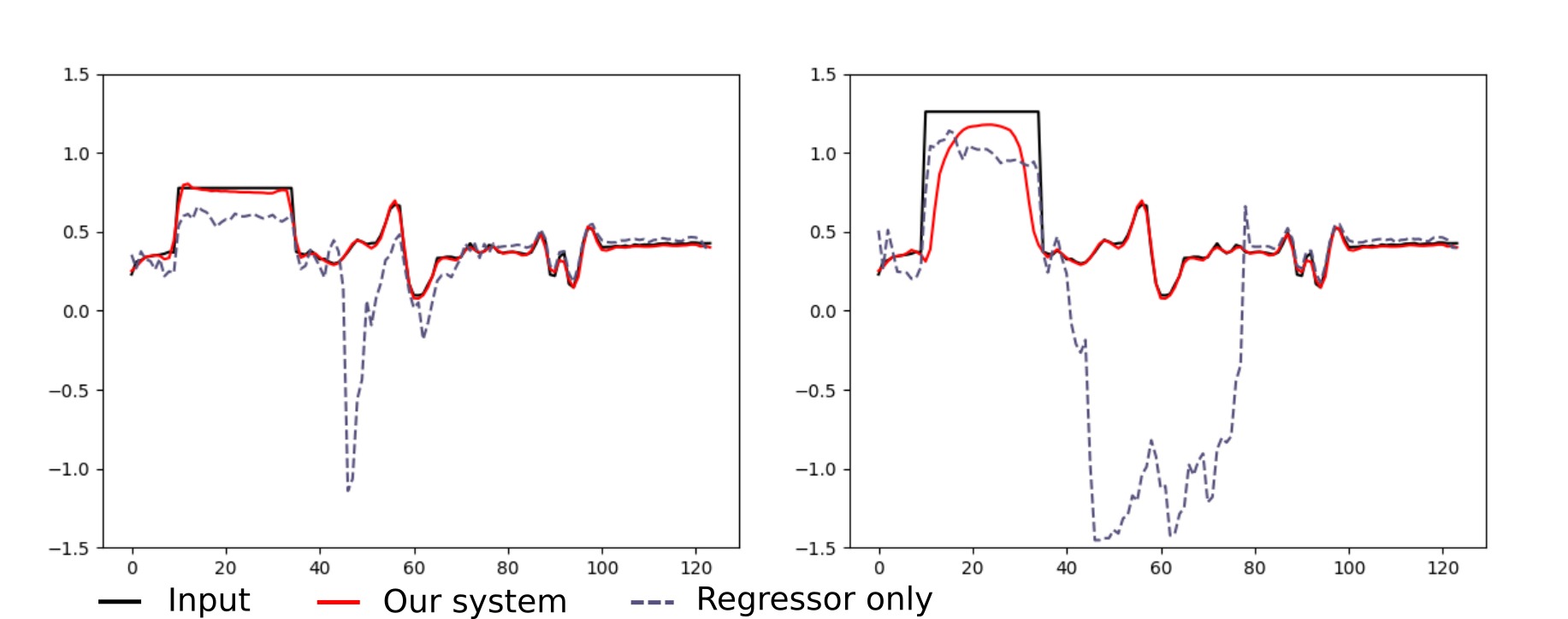}
  \caption{\
          \label{fig:open_mouth_curve}
          Realistic (left) and unrealistic (right) mouth opening input signal and the corresponding output with our system with and without the autoencoder.
          We can observe that the regressor alone is too sensitive to the input : unrealistic patterns appear as soon as a unseen input is given. 
  }
  \Description{Realistic (left) and unrealistic (right) mouth opening input signal and the corresponding output with our system with and without the autoencoder.
  We can observe that the regressor alone is too sensitive to the input : unrealistic patterns appear as soon as a unseen input is given.} 
\end{figure}

\begin{figure}[h]
  \centering
  \includegraphics[width=.45\textwidth]{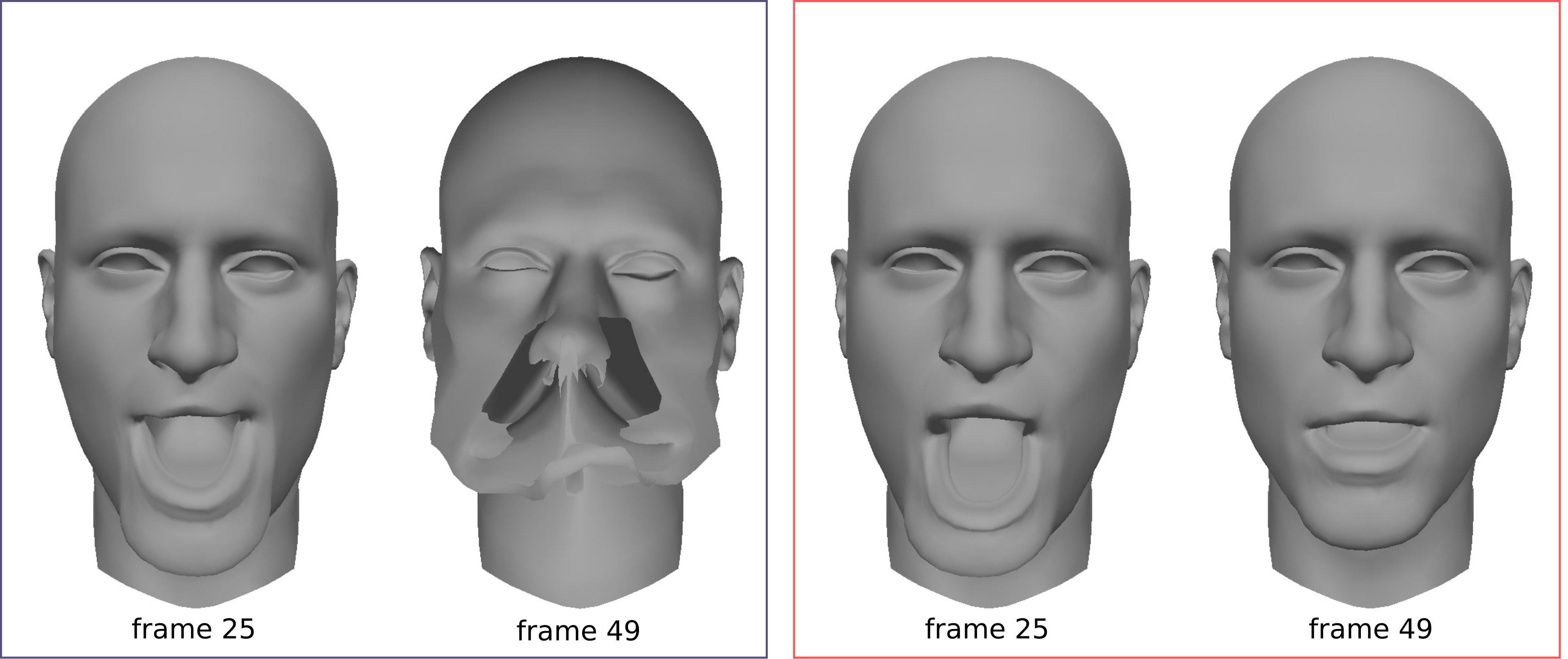}
  \caption{\
          \label{fig:open_mouth}
          Output animation with an unrealistic mouth opening without (left) and with the autoencoder (right).}
  \Description{Output animation with an unrealistic mouth opening without (left) and with the autoencoder (right).}
\end{figure}

Here we evaluate the robustness of our system by its ability to handle inadequate input.
It shows that using the regressor alone would be more accurate than the full system in term of MSE as shown in Table~\ref{tab:quantitative_res}.
However, without the autoencoder, the regressor alone would be too sensitive to user's inputs, leading to unrealistic animation output as soon as
input control parameter did not match a realistic animation.
The regressor handles the accuracy of mapping from control parameters to blendshapes animation, while the subsequent autoencoder keeps the resulting animation inside the space of plausible animation. Both components are essential for a system aimed at non-specialist users.
We show this by inputing different mouth-opening constraints and looking at inner-lips distance at output, as curves on Figure~\ref{fig:open_mouth_curve} and visually on Figure~\ref{fig:open_mouth}. 
We can see that the regressor is unstable; as soon as the input constraints constitute an unrealistic facial pattern, the output shapes are unrealistic.
The autoencoder cleans up the output animation of the regressor, generating a natural animation. 
For instance, it projects unrealistic mouth openings to realistic ones when it is required. Note that this is not just a geometric projection operation but a temporal one as well, as our autoencoder models time-windows of animation.
More results on full animations are provided in the supplementary video.

\subsection{Usability: integration in a traditional facial animation pipeline}
\label{sec:qualitative_res}
Even if our system processes whole sequences of animation, its architecture is light and performs network inference very quickly. This renders interactive uses of such a system imaginable.
In this work, we propose an interactive editing tool that is meant to be easily integrated in a facial animation pipeline that would enable non-specialist users to generate quality facial animation.
A common modern performance-based facial animation pipeline consists in acquiring sequences of actor performance, tracking his/her facial expressions, retargeting those to blendshapes animation coefficients, and finally manually tuning the obtained animation.
Today, real-time face tracking methods enable non-expert to get raw facial animation from simple video feeds, but the animation is often noisy. Moreover, as in professional pipelines the animation must often be edited later on to match the artistic intent.
Our tool finds its place at the editing stage of the pipeline.
Through an interactive interface, the user can continuously refine the animation to produce the desired animation with low-latency.
Indeed, the inference time, time between the moment the user applies its new control parameters and the moment the new final animation is produced is in average less than 0.015s for a typical scene of 8 seconds (202 frames) on CPU.\\

\begin{figure}[h]
  \centering
  \includegraphics[width=.45\textwidth]{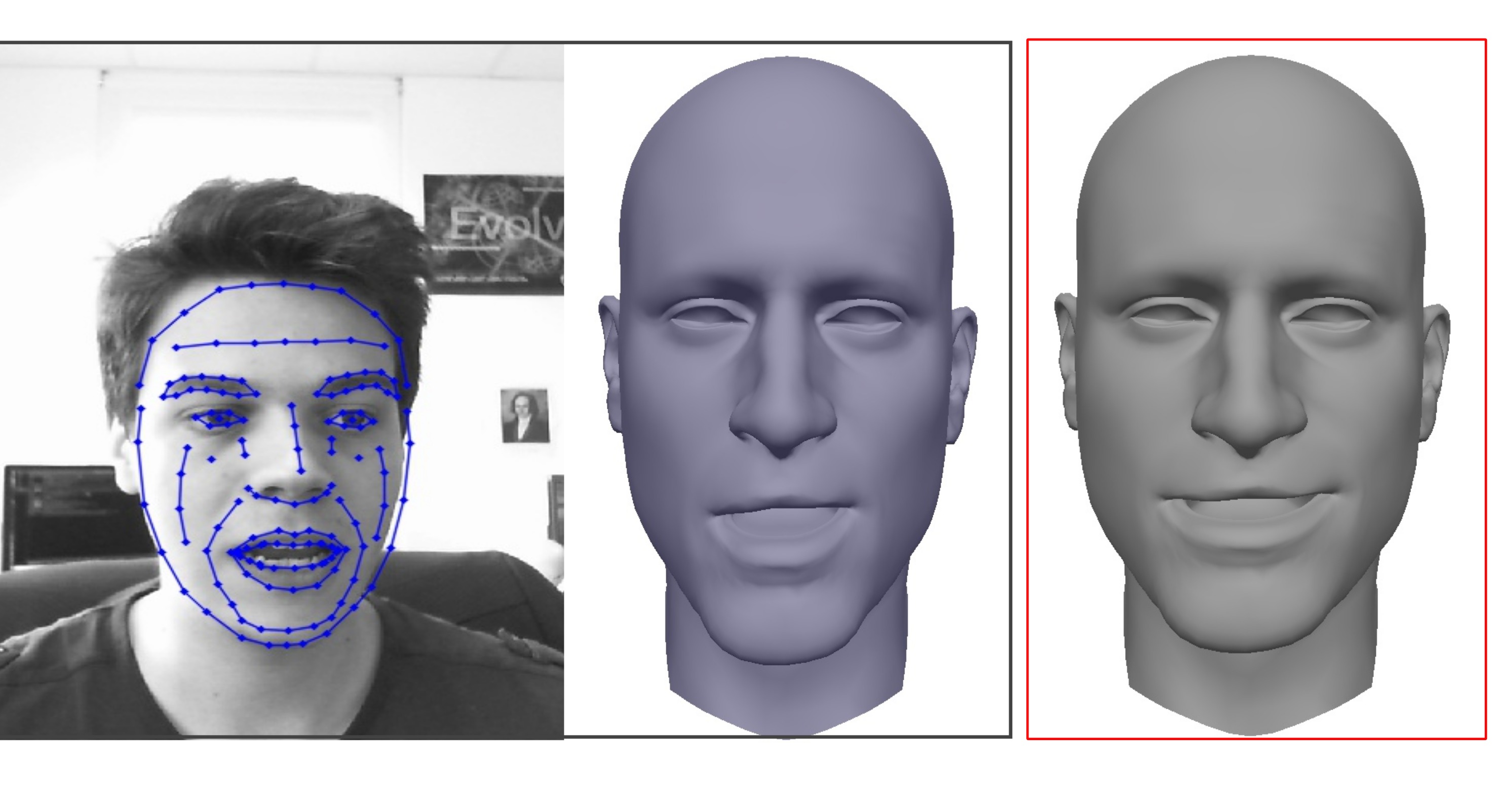}
  \caption{\
          \label{fig:from_tracking}
          Edition of a video-based animation: example of frame with 2D tracking landmarks, the animation given by the tracking sofware and the final edited animation. 
  }
  \Description{Edition of a video-based animation: example of frame with 2D tracking landmarks, the animation given by the tracking sofware and the final edited animation.} 
\end{figure}

To showcase this, we use an off-the-shelf real-time face tracking software that outputs blendshapes coefficients.
We developed a user interface that enables to visualize temporal curves for our control parameters and edit them via click-and-drag.
Our network then runs inference to deliver the edited facial animation at interactive rate. One can for instance change a neutral speech animation sequence by increasing the mouth corners distance, causing the character to smile while speaking.
Figure~\ref{fig:from_tracking} shows a frame with the 2D tracking landmarks, the corresponding animation given by the tracking as well as the final edited animation with a smile.
More isolated edits can be performed such as forcing a mouth closure or a blink by acting on the relevant local frames. 
Dynamic results of such edits are presented in the supplementary video.

\begin{figure*}[h]
  \centering
  \includegraphics[width=.95\textwidth]{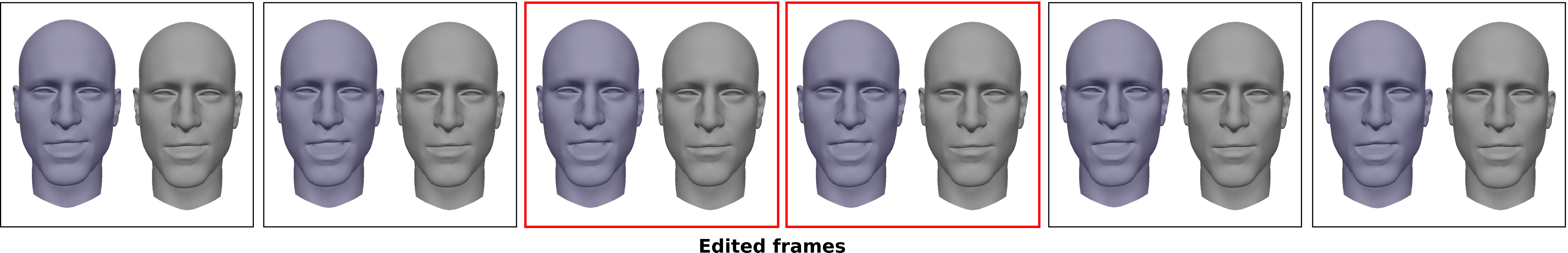}
  \caption{\
          \label{fig:close_mouth}
          Few frame edition : enforce the mouth closure.
  }
  \Description{Few frame edition : enforce the mouth closure.}
\end{figure*}

\section{Conclusion \& future work}

In this paper, we have presented a learning-based editing system that enables
easy manipulation of facial animation with simple and intuitive control parameters.
This tool can be used by non-specialist users to complete their facial animation pipeline with a tool that can correct and alter animation with no experience in facial animation.
Our method is content-independent and emphasizes robustness, resulting in an editing tool that outputs plausible animation even when given unprecise or unrealistic user inputs.
We have studied our systems behavior by evaluating quantitatively on error and semantic metrics versus relevant previous work, and have experimented with different datasets.
We have demonstrated the necessity of using resolution-preserving architecture neural network to retain 
the temporal high-frequency information of facial motion, which architectures from previous work did not address. 
To be able to train our system and perform quantitative and qualitative evaluation we have reprocessed and augmented the dataset of B3D(AC)\^{}2, and we plan to make this data available for reproducibility.

\begin{figure}[h]
  \centering
  \includegraphics[width=.35\textwidth]{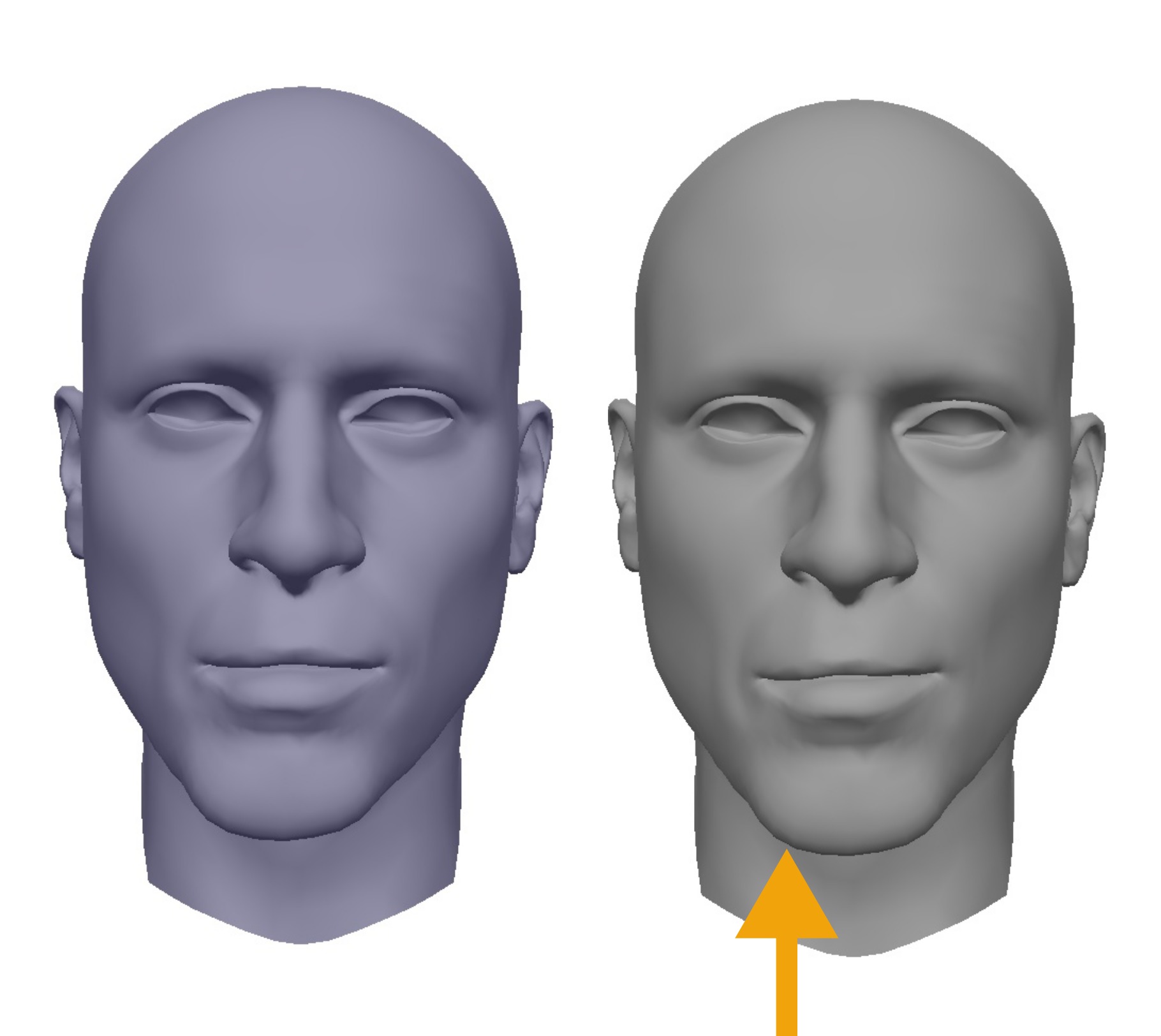}
  \caption{\
          \label{fig:limitation_chin_motion}
          Limitation of our system: some motion such as the sidewise motion of the chin of the groundtruth (left) is lost at the output of our system (right).}
          \Description{ Limitation of our system: some motion such as the sidewise motion of the chin of the groundtruth (left) is lost at the output of our system (right).}
\end{figure}

One important main limitation comes from the quality of this dataset.
Indeed, the native capture frame rate of the videos is 25 fps, which is too low to acquire all relevant natural facial cues. Important high-frequency information has already been lost at acquisition time.
We also note that the performance of our system strongly depends on the choice of the control parameters. More parameters result in a more accurate but less intuitive system that is harder to manipulate. Conversely, few parameters cause ambiguity in mapping controllers to facial shapes, resulting in less control over the produced animations.
As an example, the sidewise motion of the chin is lost due to the lack of dedicated controller (see Figure~\ref{fig:limitation_chin_motion}). 
Moreover, in our current implementation these parameters have to be continuous. 
An interesting direction of research would be to study the possibility to provide discrete inputs, even semantic ones -such as phonemes-, to control the generated animation.

\begin{acks}
We thank Emily Guy for her unnamed viewer.
We also would like to thank all the members of Dynamixyz for his software Performer.
We also thank Renaud Seguier for his supervision.
\end{acks}

\bibliographystyle{ACM-Reference-Format}
\bibliography{references.bib}

\end{document}